# Watching the acetylene-vinylidene intramolecular reaction in real time

Y.H. Jiang[1,2], A. Senftleben[1,3], A. Rudenko[4,5], M. E. Madjet[6,7], O. Vendrell[6], M. Kurka[1], K. Schnorr[1], L. Foucar[4,8], M. Kübel[9,10], O. Herrwerth[9], M. Lezius[9], M.F. Kling[9,10], J. van Tilborg[11], A. Belkacem[11], K. Ueda[12], S. Düsterer[13], R. Treusch[13], C.D. Schröter[1], R. Santra[6,14], J. Ullrich[1,4,15], and R. Moshammer[1,4]

[1]*Max-Planck-Institut für Kernphysik, 69117 Heidelberg, Germany*
[2]*Shanghai Advanced Research Institute, Chinese Academy of Sciences, 201210 Shanghai, China*
[3]*Universität Kassel, 34132 Kassel, Germany*
[4]*Max-Planck Advanced Study Group at CFEL, 22761 Hamburg, Germany*
[5]*J.R. MacDonald Laboratory, Department of Physics, Kansas State University, Manhattan KS 66506, USA*
[6]*Center for Free-Electron Laser Science, DESY, Notkestrasse 85, 22607 Hamburg, Germany*
[7]*Qatar Energy and Environment Research Institute (QEERI), Qatar Foundation, Doha 5825, Qatar*
[8]*Max-Planck-Institut für medizinische Forschung, Heidelberg, Germany*
[9]*Max-Planck-Institut für Quantenoptik, 85748 Garching, Germany*
[10]*Physics Department, Ludwigs-Maximilian-Universität, 85748 Garching, Germany*
[11]*Lawrence Berkeley National Laboratory, Berkeley, CA 94720, USA*
[12]*Institute of Multidisciplinary Research for Advanced Materials, Tohoku University, 980-8577 Sendai, Japan*
[13]*DESY, Notkestrasse 85, 22607 Hamburg, Germany*
[14]*Department of Physics, University of Hamburg, 20355 Hamburg, Germany*
[15]*Physikalisch-Technische Bundesanstalt, 38116 Braunschweig, Germany*

**Abstract:**

It is a long-standing dream of scientists to capture the ultra-fast dynamics of molecular or chemical reactions in real time and to make a molecular movie. With free-electron lasers delivering extreme ultraviolet (XUV) light at unprecedented intensities, in combination with pump-probe schemes, it is now possible to visualize structural changes on the femtosecond time scale in photo-excited molecules. In hydrocarbons the absorption of a single photon may trigger the migration of a hydrogen atom within the molecule. Here, such a reaction was filmed in acetylene molecules ($C_2H_2$) showing a partial migration of one of the protons along the carbon backbone which is consistent with dynamics calculations on *ab initio* potential energy surfaces. Our approach opens attractive perspectives and potential applications for a large variety of XUV-induced ultra-fast phenomena in molecules relevant to physics, chemistry, and biology.

Photo-induced isomerization, the migration of nuclei within a molecule upon photo excitation or ionization, often triggers primary and important biochemical reactions for instance in photosynthesis, eye vision, and viral infection (HIV isomerization)[1-3] It is at the heart of biochemical reaction dynamics often involving a sequence of transient species that change from one into another via a variety of molecular rearrangement phenomena[4]. Femtosecond spectroscopy has been applied successfully to identify intermediate states[3,5] and to understand mechanisms such as intramolecular vibrational redistribution[6]. In this approach, the pump-pulse excites the molecule, which then initiates an ultrafast atomic rearrangement. Further absorption of one or several photons during the subsequent probe-pulse enables identification of intermediate and final states. Structural information can be obtained with X-ray or electron diffraction which in principle allow atomic resolution for imaging of individual molecules in the gas-phase[7,8]. To achieve this, the usage of a large number of particles is required because of very small scattering cross-sections and, in addition, all molecules need to be oriented in space to a very high degree. Until present, this has not been accomplished in conjunction with atomic scale imaging of molecular dynamics on the femtosecond time scale. Another approach for obtaining structural information that does not rely on orientation is ionizing a single molecule with an intense femtosecond XUV pulse and applying Coulomb-explosion imaging (CEI), which allows taking snap shots of the molecular geometry at the pulse arrival-time. The power of CEI for structure determination has been demonstrated by using e.g. femtosecond laser pulses[9,11] or by sending molecular ions with MeV energies through thin foils to strip-off several electrons within a very short time thereby triggering Coulomb-explosion[10]. With commonly used infra-red lasers this requires very strong laser fields which themselves may influence the molecular structure, and foil-induced CEI is incommensurate with pump-probe schemes. Therefore, our approach here utilizes XUV light for Coulomb-explosion imaging where multiple charges can be induced by the absorption of only very few energetic photons, which hardly perturbs the geometry of the system. As the duration of the pulses has to be shorter than the investigated dynamics, and the pulse intensities need to be sufficiently high, the XUV-pump – XUV-probe spectroscopy of molecules requires the use of a free-electron laser (FEL).

Acetylene ($C_2H_2$) has been serving as a prototype for studying isomerization-induced molecular rearrangement[11-13]. Upon ionization by an XUV photon, the acetylene cation can become highly excited in the trans-bending vibrational mode, which eventually causes one hydrogen atom to migrate from the vicinity of one C atom to the vicinity of the other C atom, as illustrated in the top row of fig. 1. This involves the transformation from the linear ($D_{\infty h}$) acetylene cation to the non-linear ($C_{2v}$) vinylidene $[H_2CC]^+$ structure. Recently, employing the XUV-pump – XUV-probe approach at 38 eV we performed the first time-resolved experiment to directly observe the existence of ultra-fast isomerization in acetylene cations[14]. There, the appearance of the $CH_2^+$ fragment as a function of the pump-probe delay was used as an *indirect* indicator of the progress of the reaction. Using mixed quantum-classical dynamics calculations, Madjet et al.[15] confirmed that the first excited $A^2\Sigma_g$ state plays a key role in this ultra-fast reaction. The isomerization process leads to a region of a conical intersection with the electronic ground state of the cation close to the vinylidene structure that mediates the ultrafast electronic relaxation of the system from the $A^2\Sigma_g$ state to the ground state. Meanwhile, theoretical approaches predicted possibilities for migration recurrence via conical intersections[11,15,16] and for monitoring isomerization pathways by photoionization[17,18].

In the present work, a *direct* probing of the progress of the reaction in time is achieved by recording the momenta of the fragments after Coulomb explosion, which allows for reconstruction of the geometry at the instant of probing. A photon energy of 28 eV was chosen to avoid contribution from one-photon double ionization, which sets in at 32 eV. The experiment was performed at the Free-Electron LaSer at Hamburg (FLASH)[19] and the pump-probe pulse pairs were created by a split-and-delay mirror assembly that focuses the FEL radiation on the acetylene target (see fig.1). A reaction microscope[20] was used to resolve the fragments' momenta and to relate coincident particles with each other (see Methods section).

In a first step we discuss the reaction channel where we detect $C^+$ and $CH_2^+$ ions in coincidence, which originate from the vinylidene structure and, hence, are a tracer of isomerization. The measured yield is presented in fig. 2 as a function of the delay $\Delta t$ between the two pulses. Thereby we have only selected events where the sum kinetic energy of the two ions is larger than 5.8 eV, because these are created by ionization of a vinylidene cation by the probe pulse[14]. Therefore, the signal increases as a function of time, because isomerization has to have happened between the two pulses. Fitting the data with an exponential model function of the formula $A - B\exp(-\Delta t/\tau)$ (solid red line in fig. 2) we obtain a time constant $\tau = 65 \pm 25$ fs which we interpret as the isomerization time.

The current result is larger, but still in agreement with τ=52±15 fs extracted from our previous measurements at 38 eV[14]. In the present experiment, a systematic shift to larger isomerization times might result from the pulse duration of about 80 fs (FWHM) determined from nonlinear autocorrelation in argon[21], which is larger by a factor 2 compared with the measurements at 38 eV. However, previous measurements[22] and simulations using a partial-coherence method[23] showed that the spiky structure of the individual FEL pulses can enhance the time resolution of pump-probe experiments.

In order to reveal the momentary structure of the molecule through CEI at the time of the probe pulse one needs to measure at least three fragments in coincidence. For the observation of hydrogen migration in acetylene one should detect a proton and the two heavy ions $C^+$ and $CH^+$: Isomerization will manifest in a change of the angle between the $H^+$ and one of the heavier fragments. In order to realize this scenario the following two-pulse reaction sequence is aimed for: In the pump pulse single-photon ionization populates the $A^2\Sigma_g$ state in the acetylene cation:

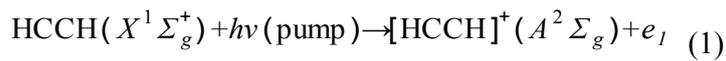

$$HCCH(X^1\Sigma_g^+) + h\nu(\text{pump}) \rightarrow [HCCH]^+(A^2\Sigma_g) + e_1 \quad (1)$$

In the probe pulse, two more photons create the final triply charged state that will quickly dissociate by Coulomb explosion:

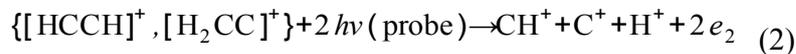

$$\{[HCCH]^+, [H_2CC]^+\} + 2h\nu(\text{probe}) \rightarrow CH^+ + C^+ + H^+ + 2e_2 \quad (2)$$

The second step is assumed to be accessible either from the acetylene cation, the vinylidene cation or any unspecified intermediate configuration. In the top row of fig. 1 the experimental scheme is illustrated on an isomerizing molecule. Note that the Coulomb explosion induced by the probe pulse produces a snapshot of the molecular geometry at the current time delay Δt. By varying Δt one can create a moving picture of the intramolecular reaction. Within the two pulses other pathways than the one described by eqns. 1 and 2 can produce $H^++C^++CH^+$ final fragments, most notably triple ionization within a single pulse. These pathways are responsible for more than 90% of the coincidences but they do not exhibit a time-dependence and therefore create a constant background to the individual pictures. To visualize the isomerization channel we have therefore subtracted the time-averaged data from the time-dependent results as discussed in the following.

We represent the three-body coincidence data in form of Newton diagrams. Thereby, we use the fact that the break-up is always planar, and plot the momenta in a two-dimensional representation. In the laboratory frame the common plane of the three fragments' momenta is arbitrarily oriented in space because we detect events for all molecular orientations. During data analysis we rotate the three momentum vectors of each event such that they have only non-zero components in the (x,y) plane. Furthermore, the vectors are rotated such that the proton has only a non-zero component in the x-direction, which in addition is always positive. This means that the proton is always directed to the right (see illustrations in the top row of fig. 3), but its momentum is not drawn. Instead, we plot the momentum distributions of the two heavy fragments in the common (x,y) plane where the y momentum components of the $CH^+$ ions are restricted to positive values and those of the $C^+$ fragments to negative values. The experimentally obtained Newton diagrams for the three time intervals of 0−60 fs (a), 40−100 fs (b) and 80−140 fs (c) are displayed in the middle row of fig. 3. Two groups of events can be identified in each diagram: In the first group the momenta of the heavy fragments are close to zero, which we interpret as molecules that have started dissociating before the final charge state has been reached. Here, Coulomb explosion imaging cannot be used to infer the molecular structure. In the second group the momenta of the $C^+$ and the $CH^+$ fragments form a ring-like structure with a radius of approximately 70 a.u. The centre of this ring is offset from the origin in x-direction by about -20 a.u., reflecting the recoil of the proton's momentum. These values correspond well with a Coulomb explosion where the ions are initially separated by typical distances in the bound molecule. Therefore, these ion triples are created either quickly within a single pulse, or the molecule is still bound after the pump pulse and the explosion is launched during the probe pulse. While the former events are time-independent and therefore subtracted from the results shown in fig. 3, the latter contain the moving picture of hydrogen migration.

To analyze the Newton diagrams we focus on the angle $\theta_{H-CH}$ spanned by the momentum vectors of the proton and the carbyne ion. Note that the angle $\theta_{H-C}$ between $H^+$ and $C^+$ contains redundant information. For overlapping pulses (fig. 3(a)), $\theta_{H-CH}$ peaks at around $145^0$, indicating a nearly linear geometry as expected in rapid triple ionization of neutral acetylene. As the time delay $\Delta t$ increases (figs. 3(b) and (c)) the carbyne ion's momentum distribution becomes broader along the $\theta_{H-CH}$ coordinate, but the centroid clearly moves towards smaller angles. At the same time, the momenta of the $C^+$ ions move in the opposite direction, away from the proton. This shows that we actually image the migrating hydrogen atom. At time delays larger than 140 fs (not shown), no changes in the Newton plots relative to fig. 3(c) are found, which is consistent with the time-dependent yield of $C^++CH_2^+$ coincidences that saturates at similar values of $\Delta t$ (fig. 2). The temporal trend in the experimental Newton plots is reproduced well by the results of our mixed quantum-classical calculations displayed in Fig. 3(d-f):

The calculations are further explained in the Methods section. Briefly, a swarm of classical trajectories for the atomic positions is started on the *ab initio* $A^2\Sigma_g$ potential energy surface of the cation close to the minimum energy geometry of the neutral system. Non-Born-Oppenheimer effects in the dynamics of the singly charged cation are described based on a surface-hopping approach[16]. For every possible time delay, the effect of the probe pulse is described by promoting the three singly charged ions onto a pure Coulombic potential and solving Newton's equations on it. It is expectable that the true interatomic potential of the triply charged system is softer than the pure Coulombic model. Indeed, in fig. 3(d-f), the momenta of $C^+$ and $CH^+$ were scaled down by a factor of 1.25 for comparison with the experimental results. It is noteworthy that the region around $\theta_{H-CH}=145^0$, which corresponds to acetylene geometry, is depopulated significantly in the theoretical results but not so in the experimental data, which may either be a consequence of the Coulombic model used to calculate the Newton maps or a contribution from other ionization pathways.

Isomerization in acetylene can occur in the cation as well as in the dication. The time-dependent result for the two-particle $C^++CH_2^+$ results could unambiguously be attributed to the cation. For the triple coincidence imaging a dication signature could also be possible, although quite unlikely, because the most significant isomerization channel through the first excited triplet state will quickly dissociate into $C^+$ and $CH_2^+$ [11, 24] and therefore does not leave a trace in the ring-like structure of the Newton diagrams. However, hydrogen migration can also lead to triplet vinylidene dications that are stable in a small region of energy transfer[24]. This reaction could contribute to time-dependencies in the Newton diagrams. On the other hand, the radius of the circle containing the $C^+$ and $CH^+$ ions points towards cation isomerization, because it corresponds to an energy coinciding with the characteristic kinetic energy release associated with hydrogen migration in the $A^2\Sigma_g$ state observed through $C^++CH_2^+$ coincidences[14,25]. An experimental route to further disentangle the two migration mechanisms would be to use significantly different intensities in the two pulses: While a stronger pump pulse enhances the dication channel, virtually only cations will contribute with a very weak pump.

In summary, a molecular movie was recorded of a hydrogen atom migrating from one carbon site to the other in the acetylene cation within less than 100 fs. The experimental results are supported by molecular dynamics calculations yielding consistent Newton maps and indicating that the main contribution to isomerization results from the $A^2\Sigma_g$ state of the acetylene cation. Our triple-coincidence Coulomb explosion imaging approach in a pump-probe setting gives structural information during the course of an intermolecular reaction, suggesting that this can be realized for other molecular systems as well. In the future, the combination with coincident electron spectroscopy is expected to provide in addition detailed information about the intermediate molecular states that are populated by the pump pulse and serve as a starting point for a specific reaction. Then, as a result of current advances in FEL technology towards higher pulse repetition rates and ever shorter pulse durations, the recording of state-resolved molecular movies with time resolutions of a few femtoseconds or even better will become feasible.

**Acknowledgements:**

The authors are greatly indebted to the scientific and technical team at FLASH, in particular the machine operators and run coordinators. Y H.J. is grateful for support from the National Basic Research Program of China (973 Program) (grant 2013CB922200), the NSFC (Grant NO. 11274232), and the Shanghai Pujiang Program (grant 13PJ1407500). M. Küb., O. H., M. L., and M.F.K. acknowledge support via the Cluster of Excellence: Munich Center for Advanced Photonics. K.U. is grateful to MEXT for support via the X-ray Free-Electron Laser Utilization Research Project and the X-ray Free-Electron Laser Priority Strategy Program. A.B. and J.v.T. were supported by CSGB/BES/DOE.

**Methods:**

Our experimental setup consists of a reaction microscope[20] equipped with an on-axis back-reflection split-mirror setup for pulse-pair creation and focusing of the FEL beam. The spherical multi-layer mirror (1 inch Mo/Si mirror, 50 cm focal length) is cut into two identical "half-mirrors". While one half-mirror is mounted at a fixed position, the other one is movable along the FEL beam axis by means of a high-precision piezo-stage, which leads to an adjustable time delay up to ±2 ps at a resolution of <1 fs. The mirror has a reflectivity of ≈30~50%, sharply peaked around the selected photon energy of 28 eV such that higher order harmonic radiation from the FEL is efficiently suppressed. During measurements the intensity of the incoming FEL beam (10 mm diameter) was equally distributed over both half-mirrors and the foci were merged inside a dilute and well localized (less than 1 mm diameter) supersonic gas-jet of acetylene molecules in the centre of the reaction microscope. With a focus diameter of <10 μm and pulse energies of a few μJ, we reached peak intensities of $I \cong 10^{11}-10^{13}$ W/cm$^2$. The spatial overlap of the two foci was initially adjusted observing an optical interference pattern obtained with a HeNe laser aligned along the beam line with ~100 μm precision. Further adjustment and fine tuning was performed with the FEL beam using non-linear $Ar^{n+}$ (n>2) ionization events. Ionic fragments were projected by means of an electric field of 47 V/cm onto a time- and position-sensitive detector (diameter 120 mm, position resolution 0.5 mm). From the measured time-of-flight (resolution 0.5 ns) and position of each individual fragment the initial momentum vectors for ions were reconstructed.

The molecular dynamics calculations were based on a mixed quantum-classical approach in which the electronic structure of the molecule is solved quantum mechanically and the dynamics of the nuclei is treated by classical mechanics. In order to describe electronic transitions close to the conical intersection between the $A^2\Sigma_g$ electronic state and the ground state of the cation we used our own implementation of the well-known Tully fewest switches algorithm[16]. An ensemble of 100 initial atomic positions and velocities were selected with a Monte Carlo procedure from a thermal ensemble in the ground electronic state of the neutral system and placed at t=0 on the $A^2\Sigma_g$ potential energy surface of the cation. The electronic structure of the cation was described at the complete active space self-consistent field level of theory with 8 correlated orbitals and 9 electrons. The forces were calculated on-the-fly and each trajectory was propagated for 400 fs. Further details on these calculations can be found in Ref. [16]. In order to simulate the probe step and obtain the Newton diagram at a particular time-delay the potential was suddenly switched to a purely Coulombic model consisting of the fragments $C^+$, $CH^+$ and $H^+$. Each trajectory was then propagated for 200 fs on the triply charged potential, which ensured that the momenta had reached their final asymptotic value.

**Author contributions:**

M.F.K., M.L., A.B., K.U., J.U., and R.M. conceived the pump-probe experiment. Y.H.J. and R.M. proposed the measurement. M.F.K., C.D.S., J.U. and R.M. supervised the experiment, which was operated by A.S., A.R., K.S., M.Kur., L.F., O.H., M.Küb., Y.H.J. J.v.T., S.D. and R.T.. The data were analyzed by Y.H.J. with support from M.Kur. and L.F.. M.E.M. and O.V. carried out the calculations, supervised by R.S.. Y.H.J., A.S. and O.V. drafted the manuscript, and all authors reviewed the manuscript and offered comments.

Captions

**Figure 1| Schematic of the experimental setup**. The split-mirror stage and ion-detection part of the reaction microscope (bottom), and sketches for mapping isomerization-induced changes of molecular geometries that occurred in acetylene cations by recording $C^++H^++CH^+$ employing the XUV-pump-XUV-probe approach (top).

**Figure 2| Yield of coincident $C^++CH_2^+$ ions as a function of delay times**. Solid blue lines represent experimental results and the solid red line is an exponential fit.

**Figure 3| Newton diagrams for $C^++H^++CH^+$ triple coincidences at three intervals of delay time**: (a,d) 0-60 fs, (b,e) 40-100 fs and (c,f) 80-140 fs. Experimental and theoretical data are displayed in (a)-(c) and (d)-(f), respectively. In each Newton plot, momentum distributions of $CH^+$ (upper half plane) and $C^+$ (lower half plane) are plotted in the reference plane where the proton momentum points to the right. Vector arrows (white) indicate momentum directions and values of the three fragments at the instant when three-body break-up takes place. The dashed arrows (red) indicate the migration direction of the proton.

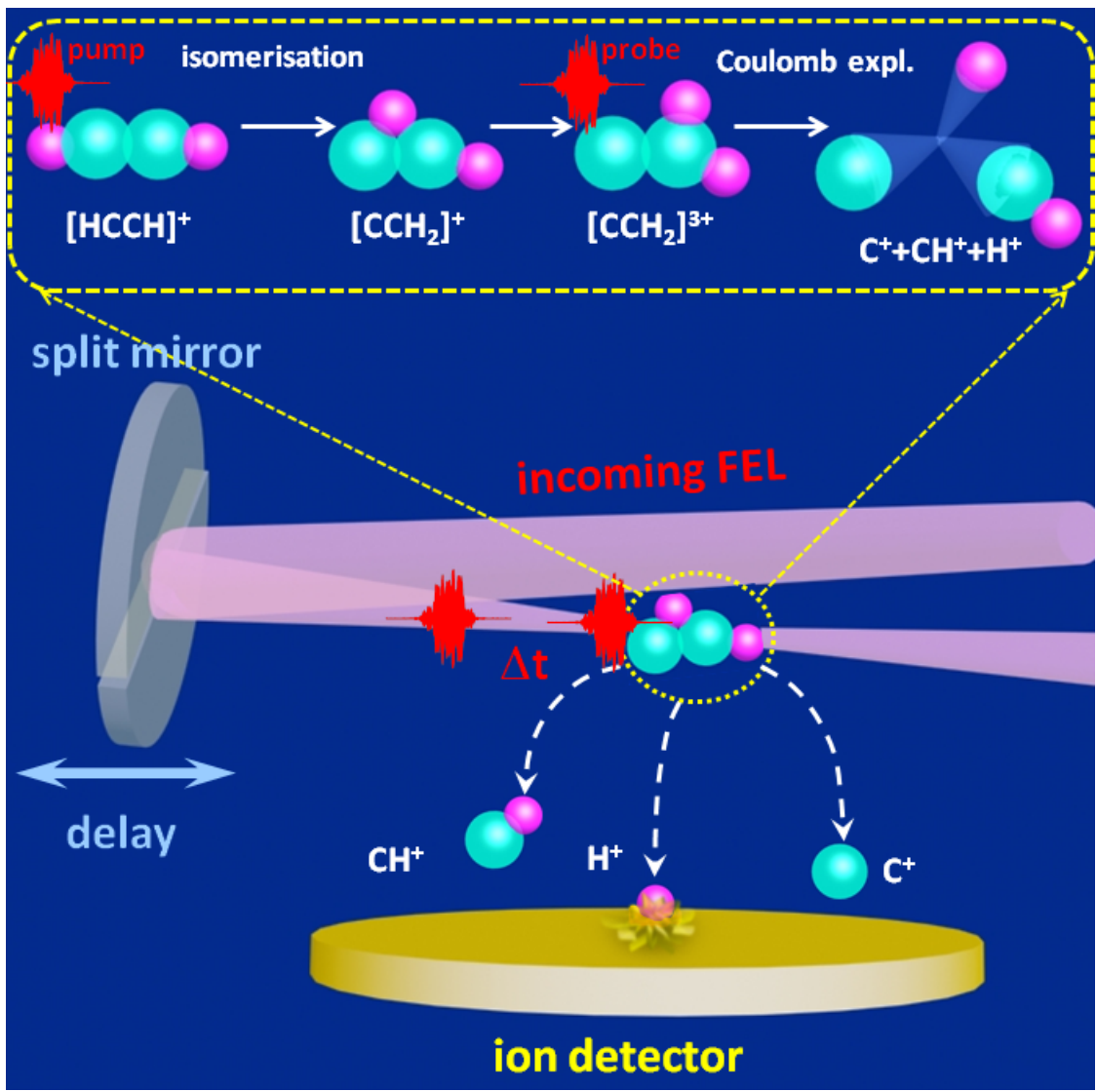

Figure 1

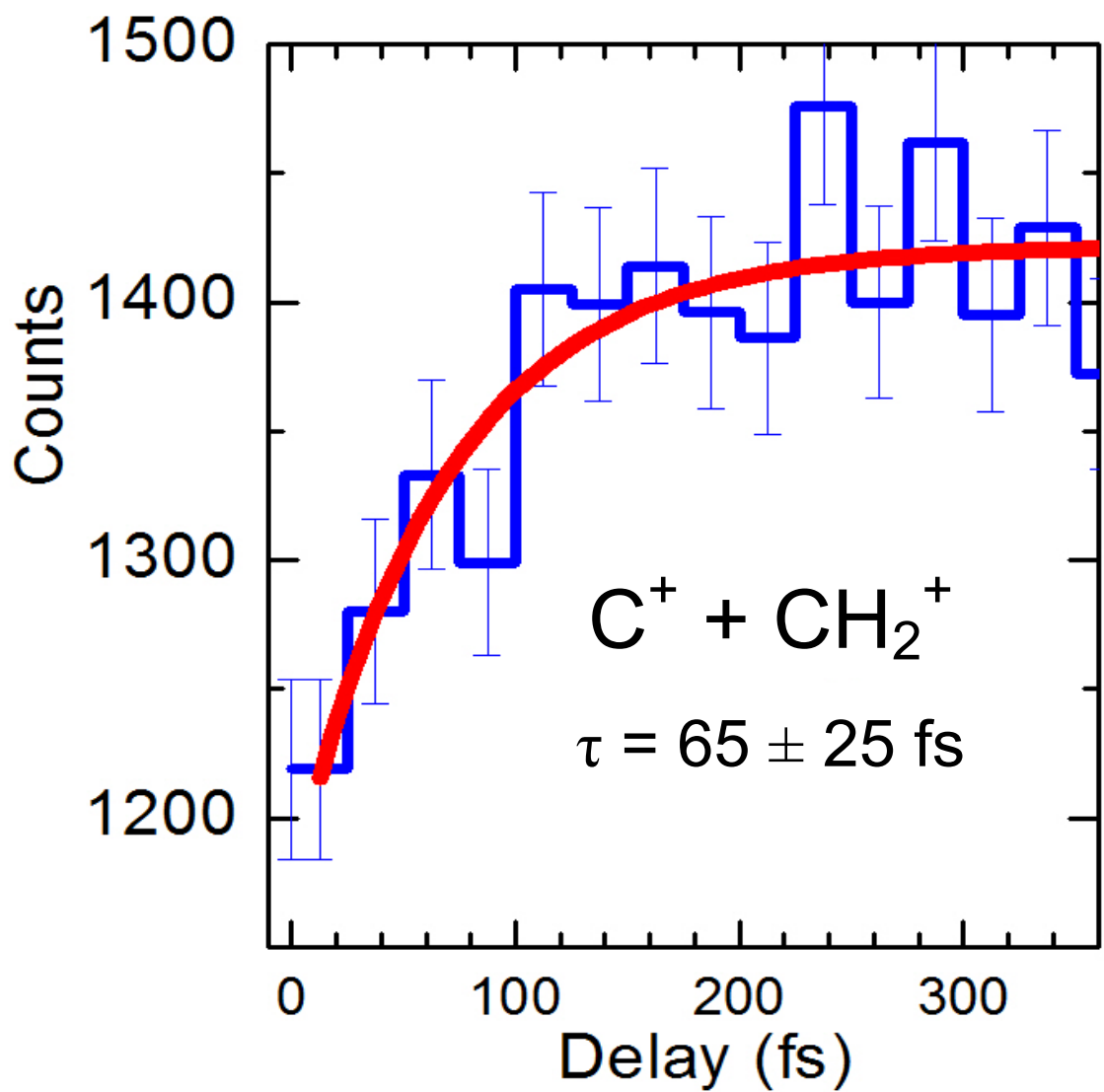

Figure 2

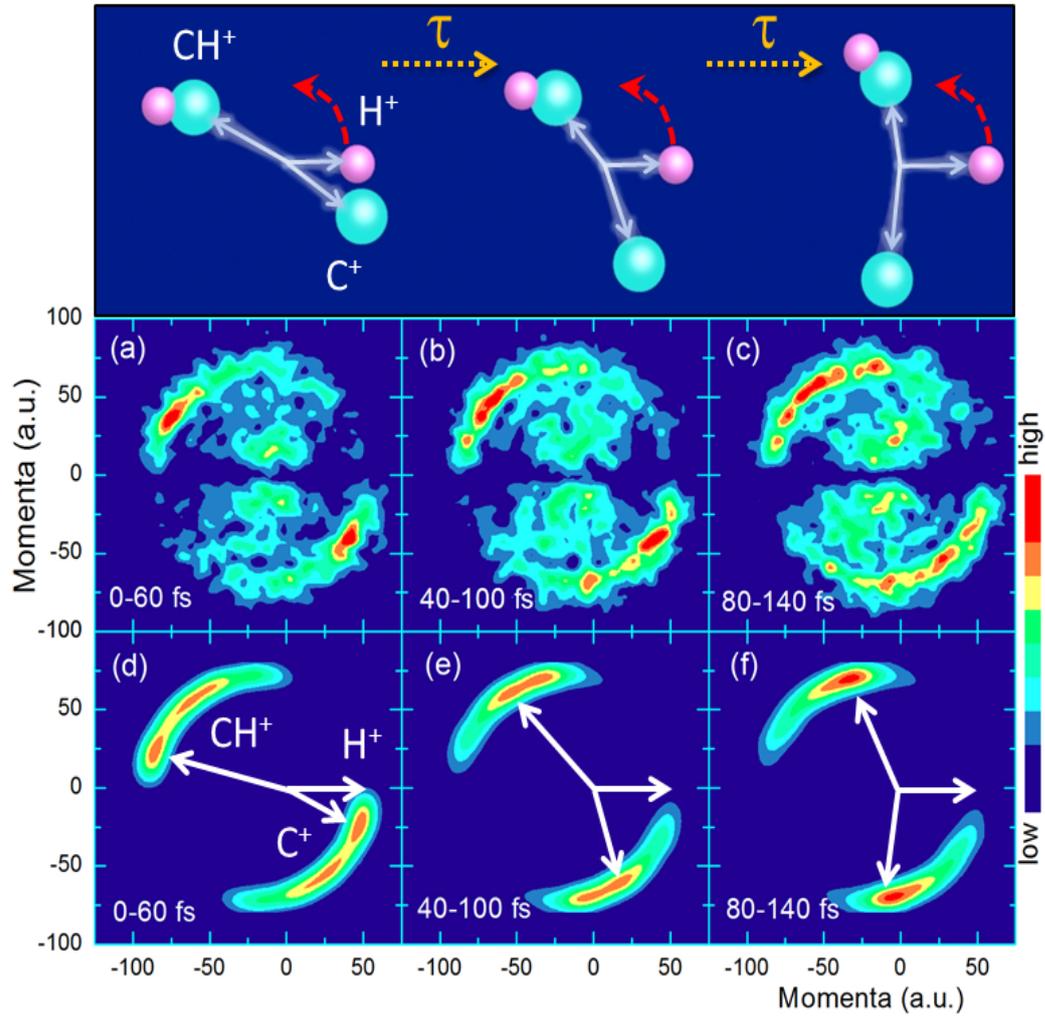

Figure 3